\documentclass[twocolumn,tighten]{aastex631}
\usepackage{multirow}
\usepackage{amssymb}
\usepackage{nicefrac, xfrac}
\usepackage{float}

\submitjournal{PSJ}

\shorttitle{Population Estimates for Theoretically Stable Centaurs}
\shortauthors{Dorsey et al.}
\graphicspath{{./}{figures/}}

\begin{document}

\title{OSSOS: XXVII. Population Estimates for Theoretically Stable Centaurs Between Uranus and Neptune}

\correspondingauthor{Rosemary C. Dorsey}
\email{rosemary.dorsey@pg.canterbury.ac.nz}
\author[0000-0002-8910-1021]{Rosemary C. Dorsey}
\affiliation{School of Physical and Chemical Sciences --- Te Kura Mat\={u}, University of Canterbury, Private Bag 4800, Christchurch 8140, New Zealand}

\author[0000-0003-3257-4490]{Michele T. Bannister}
\affiliation{School of Physical and Chemical Sciences --- Te Kura Mat\={u}, University of Canterbury, Private Bag 4800, Christchurch 8140, New Zealand}

\author[0000-0001-5368-386X]{Samantha M. Lawler}
\affiliation{Campion College and the Department of Physics, University of Regina, Regina, SK S4S 0A2, Canada}

\author[0000-0002-6722-0994]{Alex H. Parker}
\affiliation{SETI Institute, Mountain View, CA, 94043, USA}

\begin{abstract}
We calculate the upper bounds of the population of theoretically stable Centaur orbits between Uranus and Neptune.
These small bodies are on low-eccentricity, low-inclination orbits in two specific bands of semi-major axis, centred at $\sim$24.6~au and $\sim$25.6~au.
They exhibit unusually long Gyr-stable lifetimes in previously published numerical integrations, orders of magnitude longer than that of a typical Centaur.
Despite the increased breadth and depth of recent solar system surveys, no such objects have been found.
Using the Outer Solar System Origins Survey (OSSOS) survey simulator to calculate the detection efficiency for these objects in an ensemble of fully characterised surveys, we determine that a population of 72 stable Centaurs with absolute magnitude $H_{r}\leq10$ (95\%~confidence upper limit) could remain undetected.
The upcoming Legacy Survey of Space and Time (LSST) will be able to detect this entire intrinsic population due to its complete coverage of the ecliptic plane.
If detected, these objects will be interesting dynamically-accessible mission targets --- especially as comparison of the stable Centaur orbital phase space to the outcomes of several modern planetary migration simulations suggests that these objects could be close to primordial in nature.
\end{abstract}


\section{Introduction}\label{sec:intro}

Stable orbits between Uranus and Neptune were first identified and investigated in the 1990s, with several studies showing that test particles in a narrow band of semi-major axis survived tens of Myr during numerical integrations \citep{Gladman_1990, Holman_1993, Holman_1997}.
This is unusual for minor planets orbiting within the giant planet region (called `Centaurs'), as they typically occupy more dynamically excited orbits and experience short, transitional lifetimes \citep[$\sim$$10^2$--$10^4$ kyr;][]{Tiscareno_2003, Horner_2004} due to gravitational scattering during close encounters with the giant planets.

Despite their durability, no further study of the stable Centaur orbits was performed until the recent work by \cite{Zhang_2022}.
Their study offered a high-resolution characterisation of the orbital phase space shown to be numerically most stable and the first physical explanation for the sharp transitions in stability timescales in this region of the solar system.
Through the integration of $\sim$$10^{5}$ test particles in the orbital phase space between Uranus and Neptune in their current locations \citep[a factor of $\sim$$10^2$ more than that used by][]{Holman_1997}, \cite{Zhang_2022} found that dynamically cold orbits (eccentricity $e<0.05$ and ecliptic inclination $i<5^{\circ}$) in bands of semi-major axis centred at $\sim$24.6~au and $\sim$25.6~au, with widths of $\sim$0.3 and $\sim$0.4~au respectively, survived for 4.5~Gyr.
This stability timescale is three orders of magnitude longer than that of the median dynamical lifetime for inactive Centaurs \citep{Fernandez_2018} and is the same magnitude as the solar system's current lifetime.
Their analysis of the solar system resonances revealed that the Gyr orbital stability of these test particles was due to the complex spatial and temporal relationship between Uranus' and Neptune's mean motion resonances.

No stable Centaurs have been observed to date, though the simulated particles exhibit orbital stability on timescales similar to the extant populations of classical Kuiper Belt objects \citep{Lykawka_2012} and Neptune Trojans \citep{Dvorak_2007, Lin_2021}.
Recent sky surveys with wide-field imagers have quantified sensitivity to low-inclination Centaurs \citep{Kavelaars_2008, Bannister_2020}. 
For example, under the sensitivity of the Outer Solar System Origins Survey \citep[OSSOS;][]{Bannister_2016, Bannister_2018}, Centaurs are detectable down to absolute magnitude $H_{r} \lesssim 9$ at a distance of 30~au. 

In this work we quantify the maximum number of Gyr-stable Centaurs that could exist between Uranus and Neptune while remaining undetected.
The object simulation process using a survey simulator and the subsequent population estimate calculation are described in Section~\ref{sec:meth}.
We present our population upper limit for stable Centaurs in Section~\ref{sec:results}, as well as a qualitative discussion of its uncertainty.
In Section~\ref{sec:cosmo}, we compare outcomes of planetary migration simulations with the stable Centaur population and assess their similarity.
Lastly, we discuss future considerations for this population in Section~\ref{sec:future}, including our predictions for the detection and characterisation of stable Centaurs by the upcoming Legacy Survey of Space and Time (LSST).

\section{Applying the OSSOS Survey Simulator}\label{sec:meth}

\subsection{Method}
To quantify how many objects in a population have not been observed by a solar system sky survey, it is necessary to understand how efficient the survey is at detecting solar system objects on different types of orbits.
Survey characterisation (quantification of the inherent observational biases) is becoming a standard technique for modern surveys \citep{Kavelaars_2008, Bannister_2020} and is crucial for the application of survey simulators.
A survey simulator is a framework of software that simulates the detection of objects by a fully characterised survey  \citep{Lawler_2018_O10}.
It takes an input model population and applies the survey's systematic biases to produce a sample of biased detections of the model.
These simulated model detections can then be compared to real observed populations using statistical analysis.

Survey simulators are also useful for debiasing small body populations that could exist but have no detections, as is the case for the stable Centaurs.
Using Poisson statistics, the probability of obtaining zero detections when three are expected is $\sim$5\%; therefore the number of model objects required to produce three simulated detections when none are expected is the 95\% confidence upper limit on the intrinsic population size.

In this work, we adopted a different methodology for using the survey simulator for non-detections than that suggested by \citet{Lawler_2018_O10}.
Instead of running the survey simulator until three detections were simulated, we modified the OSSOS survey simulator \citep{Lawler_2018_O10} to terminate after a large number of simulated objects had been created.
This avoids a phenomenon called `survivorship bias' \citep{Dirk_2021} from affecting the generated simulated sample of objects.
Survivorship bias occurs when data is selected by passing a certain criteria and all other data is ignored.
This can lead to inaccurate conclusions, since the unused data can also contain useful information.
In the context of the survey simulator, all detections and non-detections of simulated objects that occur after the third simulated detection still contain information about the rate of detection of the population.
Terminating the survey simulator `early' (after the third detection) can lead to a preferentially higher object detection rate and therefore a lower population limit.

For the analysis of the stable Centaurs, a large sample of simulated detections and non-detections was constructed using a `blurring' process to minimise the probability of exactly replicating an object. 
We randomly offset each element of an input object's duplicated orbit by a specified fraction to produce an additional similar orbit.
The 95\% confidence upper limit on the intrinsic population size was then calculated as:
\begin{equation}
    N_{upper} = 3\times\frac{1}{f_{detection}} = 3\left(\frac{N_{sample}}{N_{detections}}\right)
\end{equation}
where $f_{detection}$ is the mean detection rate within the simulated sample, which is given by the ratio of the number of detections ($N_{detections}$) to the total number of simulated objects ($N_{sample}$).
This assumes that the sample statistics are representative of the intrinsic population.
Additionally, the survey simulator was run until the population size upper limit error was reduced to within one object.
This required $10^7$ simulated objects, as the detection rate error is inversely proportional to the sample size, $\sigma_{detection}\propto {N_{sample}}^{-\sfrac{1}{2}}$, and the population size error is given by:
\begin{equation}
    \sigma_{N_{upper}}\propto\frac{\sigma_{detection}}{{f_{detection}}^2}\propto\frac{1}{{f_{detection}}^2~\sqrt{N_{sample}}}
\end{equation}

\subsection{Input Models and Parameter Constraints}

\begin{figure*}
    \includegraphics[width=\textwidth]{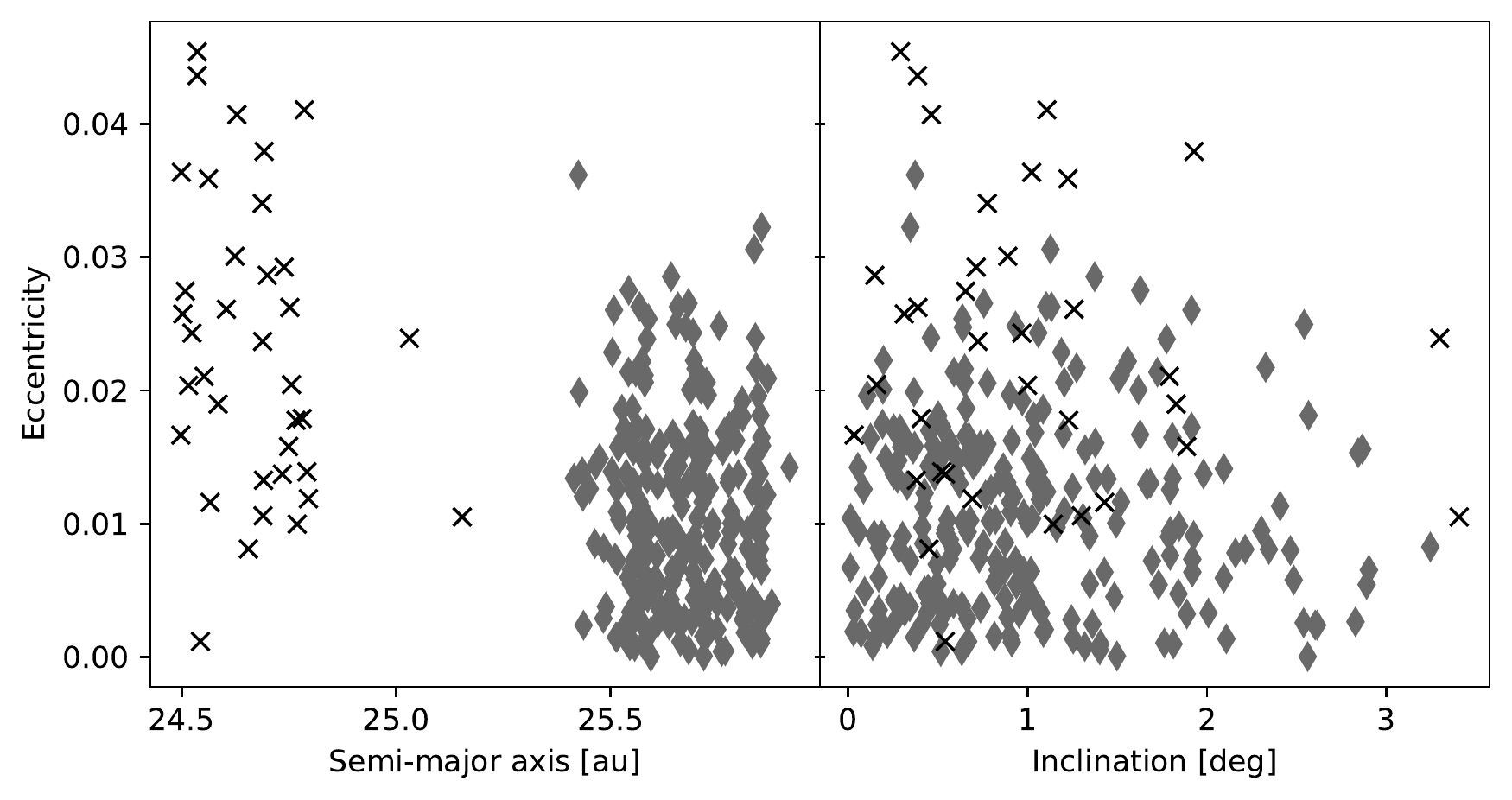}
    \caption{Orbital phase space of both the initial and final states of the test particles that survived 4.5~Gyr of integration by \citet{Zhang_2022}, used here as the input model for the OSSOS survey simulator to assess the size of the intrinsic population. The crosses and diamonds indicate states of the inner band ($a<25.2$~au) and the outer band ($a>25.2$~au), respectively}. For a comparison between the initial and final states of individual particles, see Fig. 6 of \citet{Zhang_2022}.
    \label{fig:Gladman}
\end{figure*}

We used the \textit{r}-band efficiency and detections of three characterised outer solar system surveys that used the Canada-France-Hawaii Telescope (CFHT): the Canada-France Ecliptic Plane Survey \citep[CFEPS;][]{Petit_2011}, \citet{Alexandersen_2016} and OSSOS \citep{Bannister_2016, Bannister_2018}.\footnote{While CFEPS Hi-Lat \citep{Petit_2017} is often used together with these surveys for population estimates, it does not have sky coverage of any part of the orbital phase space we consider here, so it is omitted.}
The input model used was the set of test particles that survived the integration of \cite{Zhang_2022}, using both their initial and final states in order to increase the input orbit sample size (see Figure~\ref{fig:Gladman} and Table~\ref{tab:stats}).
Since only the semi-major axis, eccentricity and inclination of these particles were recorded and these are non-resonant objects, the remaining orbital angles (longitude of ascending node $\Omega$, argument of pericenter $\omega$, and mean anomaly $M$) were drawn randomly on $[0,2\pi]$.
Each orbital element was blurred randomly by $\leq1\%$, as the input population is entirely theoretical, and therefore any uncertainties associated with the result will be larger than the uncertainty introduced in this step.

\begin{deluxetable*}{l|ccc|ccc}
\tablenum{1}
\tablecaption{Median, 95\textsuperscript{th} and 99\textsuperscript{th} percentiles for the initial and final states of the test particles from \citet{Zhang_2022}.} \label{tab:stats}
\tablewidth{0pt}
\tablehead{
& \multicolumn{3}{c|}{Initial} & \multicolumn{3}{c}{Final} \\
Parameter & Median & 95\textsuperscript{th} Percentile & 99\textsuperscript{th} Percentile & Median & 95\textsuperscript{th} Percentile & 99\textsuperscript{th} Percentile}
\startdata
Eccentricity      & 0.00884 & 0.0274 & 0.0362 & 0.0134 & 0.0285 & 0.0409\\
Inclination (deg) & 0.695   & 2.46   & 3.10   & 0.943  & 2.48   & 2.89\\
\enddata
\end{deluxetable*}

In order to assess detectability, object apparent magnitude is required, which depends on geocentric distance at the current epoch (calculated from the particle's orbit) and the population's absolute $H_{r}$-magnitude distribution, assuming no cometary activity. This is observationally consistent with the Centaurs detected in OSSOS, which were found by \citet{Cabral_2019} to display no activity.
According to \cite{Lawler_2018_O8}, neither a knee nor divot shape are rejectable for the intrinsic $H_{r}$-magnitude distribution of the observed Centaur population.
We assess their preferred parameters for each $H_{r}$-magnitude distribution type (Table~\ref{tab:results}).
The $H_{r}$-magnitude range explored in this work is $6\leq H_{r}\leq 10$, which corresponds to object diameters of $60\lesssim D \lesssim400$~km, assuming an albedo of 0.04 \citep{Duffard_2014}.
This fully encompasses the current observational constraints on Centaurs in the solar system, from the largest known Centaur \citep[Chariklo with $H_{r} \sim6.8$;][]{Peixinho_2015} to near the OSSOS observational limit for Centaurs at a heliocentric distance of $\sim$$26$~au.

\begin{deluxetable*}{lccccccc}
    \tablenum{2}
    \tablecaption{Absolute magnitude distributions and population estimates for stable Centaurs.\label{tab:results}}
    \tablewidth{10pt}
    \tablehead{
    & \multicolumn{6}{c}{Parameters} \\
    \colhead{Distribution type} &
    \colhead{$\alpha_{b}$} &
    \colhead{$H_{b}$} &
    \colhead{c} &
    \colhead{$\alpha_{f}$} &
    \colhead{$H_{min}$} &
    \colhead{$H_{max}$} &
    \colhead{$N_{upper}$}
    }
    \startdata
    Divot & 0.9 & 8.3 & 3.2 & 0.5 & 6 & 10 & $<72$ \\
    Knee & 0.9 & 7.7 & 1.0 & 0.4 & 6 & 10 & $<71$ \\
    \enddata
    \tablecomments{Column headings from $\alpha_b$ to $N_{upper}$: absolute magnitude distribution bright slope, break magnitude, contrast, faint slope, absolute magnitude lower bound, absolute magnitude upper bound, and the resulting population upper limit estimate.}
\end{deluxetable*}

\section{Population Size Upper Limit for Stable Centaurs} \label{sec:results}
We find that the divot and knee $H_{r}$-magnitude distributions produce nearly the same 95\% confidence upper limits of 72 and 71 stable Centaurs respectively with $H_{r}\leq 10$, corresponding to $D \gtrsim 50$~km (Table~\ref{tab:results}).
\cite{Lawler_2018_O8} estimated that the total population of Centaurs with $H_{r}\leq10$ is $390^{+200}_{-150}$ and $550^{+340}_{-290}$ objects, respectively.
Our estimate of $\sim$72 stable Centaurs for both distributions could therefore contribute up to $\sim$13--18$\%$ of the total Centaur population with $H_{r}\leq10$. 
This is a substantial number of objects to remain undetected if they exist.

Our results are relatively unaffected by the integration's population ratio between stable bands.
Since our analysis is dependent on the outcome of \cite{Zhang_2022}'s integration, we cannot generalise our result to all possible numerically derived instances of the theoretical stable Centaur population.
However, due to their near perfectly circular low-inclination orbits, the detectability of these stable Centaurs in an equatorial survey near opposition is primarily determined by their absolute magnitude and which of the two stable bands they occupy.
The apparent brightness of stable Centaurs with similar absolute magnitude decreases by $<0.3$~mag when moved from the closer band ($\sim$$24.6$~au) to the distant band ($\sim$$25.6$~au).
Coupling this to the assumed power-law luminosity function slopes, a population residing entirely in the outer band could be only $\sim$0.5\% larger than our stated upper limits which assume a relative occupation ratio of 1:7 between the inner and outer bands, respectively, as was the case for the initial and final states of the \citet{Zhang_2022} integration.

\section{Exploring Stable Centaur Origins From Cosmogonic Solar System Models} \label{sec:cosmo}

In Section~\ref{sec:results}, we demonstrated that a significant number of stable Centaurs could be orbiting in today's solar system while remaining undetected by surveys. This result raises further questions about the theoretical stable Centaur population, including how such orbits could have been produced by plausible planet migration scenarios.

Earlier studies have suggested it is unlikely for stable Centaurs between Uranus and Neptune to have survived in-situ from the early solar system.
Initial exploration by \cite{Brunini_1998} with only 100 test particles found that none survived on the Gyr timescales observed in previous integrations \citep{Gladman_1990, Holman_1993, Holman_1997} when additional factors such as Neptune's migration, Pluto-sized planetesimals, and mutual gravitational interactions were included.
They concluded that it is highly unlikely for a ``substantial" number of objects to exist in the region but did not quantify any upper limit.

However, more recent planetary migration simulations \citep[][hereafter referred to as the Nesvorn\`y et al. simulations]{Nesvorny_2012, Nesvorny_2013, Nesvorny_2020} with three orders of magnitude higher resolution produced multiple test particles within the (\textit{a},\textit{e}) phase space of the stable Centaurs.
These simulations have been confirmed to match the broad structure of the current trans-Neptunian region \citep{Nesvorny_2020}.
They model Neptune's migration through a primordial Kuiper Belt with slightly different parameterisations:
\begin{itemize}
    \item Simulation 1: Neptune starts at 24~au and migrates outward due to analytic terms through a disk of $10^6$ particles initialised between 24 and 30~au \citep{Nesvorny_2020}.
    \item Simulation 2: Similar to simulation 1, except with a faster migration of Neptune \citep{Nesvorny_2020}. The e-folding exponential migration timescales were $\tau_1=10$ and $\tau_2=30$~Myr compared to 30 and 100~Myr respectively for simulation 1.
    \item Simulation 3: Similar to simulation 1, except that planets feel the gravity of planetesimals instead of being forced by analytic terms \citep{Nesvorny_2012, Nesvorny_2013}. The disk of planetesimals contained $6\times10^5$ particles.
\end{itemize}

Collectively, a subset of the Nesvorn\`y et al. test particles' orbital parameters appear similar to those of the \citet{Zhang_2022} higher eccentricity or inclination stable Centaurs.
To quantify this similarity, we considered a one-dimensional parameterisation of the (\textit{e},\textit{i})~phase space using an `excitation' parameter defined as:
\begin{equation}
\label{eq:excitation}
    Excitation = \sqrt{e^2 + \textnormal{sin}^{2}i}
\end{equation}
Since eccentricity and inclination both impact an object's orbital excitation, they appear in independent terms within the excitation parameter definition so that their effect is not omitted if the other parameter is zero.

\begin{figure*}[]
    \includegraphics[width=\textwidth]{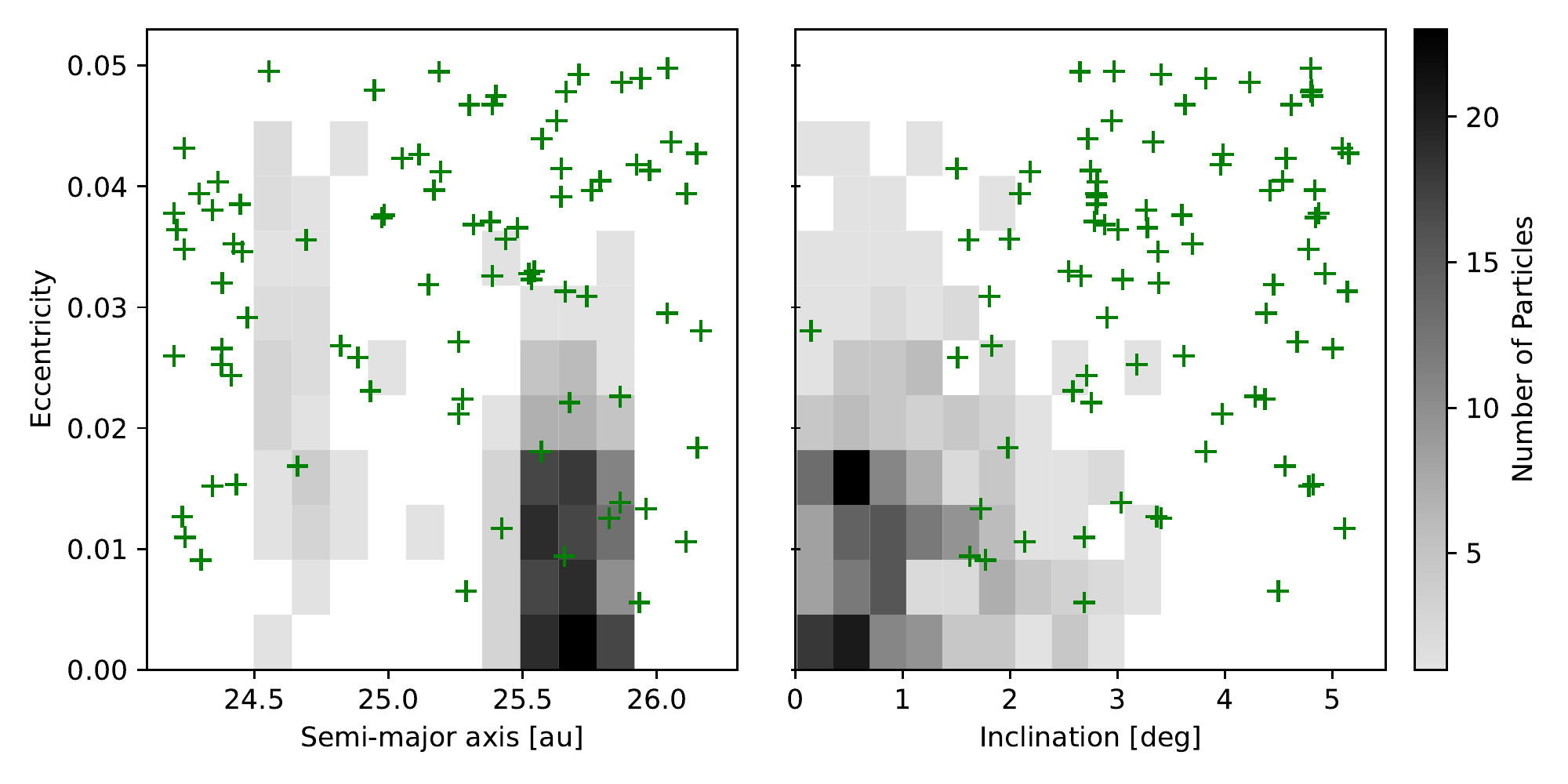}
    \caption{Overlap of the orbital phase space of the test particles that survived Gyr integration by \citet{Zhang_2022} (black density plot) and the combined test particle outputs from three planetary migration simulations (green crosses; Nesvorn\'y et al. simulations).}
    \label{fig:hexbin}
\end{figure*}

From the Nesvorn\`y et al. simulations, 3, 71 and 8 test particles were found with $24.2<a<26.2$~au and $e<0.05$ at $t=500$~Myr, respectively.
Since simulations~1~\&~3 produced only a few test particles, they do not provide a meaningful sample size for statistical comparison as individual datasets.
With the aim of merging all three simulation outputs into one dataset, we compared the excitation values of the test particles from simulations 1 \& 3 to simulation 2 using an Anderson-Darling test \citep{Anderson_1952}.
We found that neither simulation 1~or~3 are rejectable as having been drawn from the same intrinsic population as simulation~2.
Therefore, for the subsequent qualitative analysis we considered all test particles from the three simulations as a single dataset, for comparison with the \citet{Zhang_2022} particles (Figure~\ref{fig:hexbin}).

\begin{figure}[h]
    \centering
    \includegraphics[width=\columnwidth]{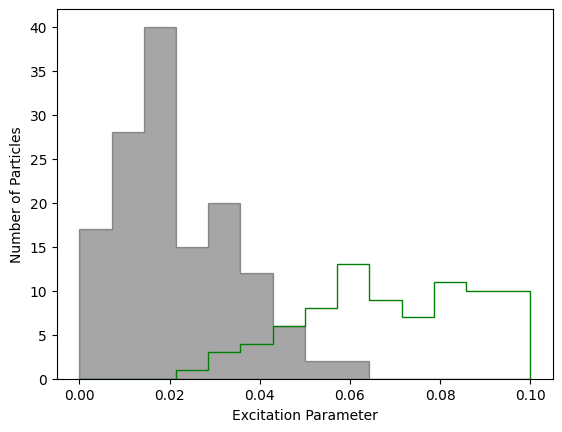}
    \caption{Excitation parameter values (Eq.~\ref{eq:excitation}) for the initial states of the \citet{Zhang_2022} stable Centaur particles (grey) and the combined particle outputs from three planetary migration simulations (green; Nesvorn\`y et al. simulations).}
    \label{fig:excitation}
\end{figure}

Considering the excitation distributions of the initial states of the stable Centaurs and the planetary migration test particles, there is a reasonable overlap (Figure~\ref{fig:excitation}).
Approximately 13\% of the stable Centaur excitation space (the grey upper tail in Fig.~\ref{fig:excitation}) is consistent with the lower tail of the planetary migration test particle excitation distribution.
This suggests that planetary migration simulations can successfully emplace test particles into the stable Centaur phase space, including more physically accurate scenarios, like simulation 3, that assign mass to planetesimals.
Planetary migration simulations are not always targeted towards replicating the solar system Centaur population, so there are definite opportunities to further probe possible origins of the stable Centaurs.
A coherent simulation of the stable population from emplacement in the early solar system formation through to the present day would also be beneficial, to confirm that the current simulations of the stable Centaur formation and evolution are fully compatible.

\section{Future observations} \label{sec:future}

The upcoming Vera Rubin Observatory's Legacy Survey of Space and Time (LSST) will be revolutionary in its discovery of solar system objects, including Centaurs, and potentially stable Centaurs.
Despite the multi-metre difference in aperture between CFHT and Rubin's Simonyi Survey Telescope, LSST's single-exposure limiting apparent magnitude of $m_r\sim24.5$ \citep{LSSTSC_2009} is comparable to or slightly shallower than that of most of the OSSOS observing blocks \citep{Bannister_2018}.
LSST will therefore be sensitive to similar $H_{r}$ limits to those we discuss here, i.e. $\sim70$ stable Centaur detections. 
Just as for OSSOS, LSST's Year 1 set of discoveries is therefore unlikely to differentiate between the current $H_{r}$ models for Centaurs, since objects with $H_{r}\geq10$ at a heliocentric distance of $r_{h}\sim25$~au will be fainter than the single-exposure limit.
Additional shift-and-stack surveys made on acquired LSST data will reach deeper, as will current but smaller-area invariant plane surveys such as DEEP \citep{Bernardinelli_2023}, FOSSIL \citep{Chang_2021}, and CLASSY \citep{Fraser_2022}.

The capability of LSST to discover and characterise solar system objects was recently evaluated by the solar system science community \citep{Schwamb_2023}.
The orbital model used in their simulations was the Pan-STARRS Synthetic Solar System Model \citep[S3M;][]{Grav_2011} which includes dynamically excited Centaurs but no stable Centaur orbits.
However, the evaluation uses the interpolation between main-belt asteroids and trans-Neptunian objects to quantify LSST's performance for the Centaur region.

While LSST's cadence optimisation process has not explicitly addressed stable Centaurs, the survey will encompass the full on-sky spread of the entire intrinsic population we describe here.
As devised in the Phase 2 Survey Cadence Optimization Committee Report, the survey will provide complete coverage of the sky within $\pm5^\circ$ of the ecliptic plane totaling 1,800~deg$^2$ \citep{ROSCOC_2023}. 
For comparison, OSSOS only covered $\sim$110~deg$^2$ of this area; $\sim$6\% of LSST's coverage.
By our population estimates and interpolation of the absolute magnitude distributions, LSST should thus be able to detect $\sim$20 stable Centaurs with $D\gtrsim100$~km.

The exact location of origin of stable Centaurs in the pre-migration disk is unclear.
The Nesvorn\`y et al. simulations imply that stable Centaurs are less likely to be \textit{in situ} survivors of Neptune's outward migration \citep[in contrast to e.g. the cold classical Kuiper Belt objects;][]{Morbidelli_2020, Gomes_2021}, as Neptune begins at 24~au and therefore crosses their orbits.
However, as these simulations model Neptune's migration as relatively ``smooth" in its changes in semi-major axis, future investigations modelling Neptune's ``grainy" migration, with stochastic changes in semi-major axis due to dwarf planet scattering, could potentially leave small pockets of objects \textit{in situ}.
The largest mass of simulated planetesimals is at 24~au, potentially providing abundant objects to capture into 24.5--26~au.
Alternatively, stable Centaurs could potentially be rarely populated by dynamically excited objects that formed nearby that are occasionally cycled into the stability region.
In either case, expected stable Centaur colours in LSST, given a likely origin inside of 30~au, would be $g-i<1.2$, akin to the more dynamically heated Neptune Trojans \citep{Nesvorny_2020}.

If found by LSST, the stable Centaur population will make intriguing future mission targets.
The stable Centaurs orbit within $\pm5^\circ$ of the ecliptic plane, making them easily accessible mission targets even under delta-v restrictions.
Alternatively, in the case of zero detections, LSST will rule out the class of Neptune migration models that populate the stable Centaur region.

\section{Conclusion}

In this work, we have further characterised the theoretical stable Centaur population described by \citet{Zhang_2022}, which occupy bands of semi-major axis centred at $\sim$24.6~au and $\sim$25.6~au with low-eccentricity ($e<0.05$) and low-inclination ($i<5^{\circ}$) 
from the ecliptic plane.
Here, we summarize our key findings:
\begin{itemize}
    \item Using the OSSOS survey simulator, we find the 95\% confidence upper limit population size to be 72 objects with $H_r < 10$.
    \item Cosmogonic simulations with various Neptune migrations show promise at populating this region at 500 Myr.
    \item LSST will be able to see this intrinsic population to near completion at $H_r < 10$ if they are present in today's solar system.
\end{itemize}

\begin{acknowledgments}
We thank Brett Gladman for helpful feedback on the manuscript, Kevin Zhang for providing their test particle data, and David Nesvorn\'y for providing planetary migration outcomes.

RCD acknowledges support from the University of Canterbury Doctoral Scholarship.
MTB appreciates support by the Rutherford Discovery Fellowships from New Zealand Government funding, administered by the Royal Society Te Ap\={a}rangi.
\end{acknowledgments}

\vspace{5mm}

\software{OSSOS survey simulator \citep[][see \url{https://github.com/OSSOS/SurveySimulator}]{Lawler_2018_O10},
Anaconda \citep{anaconda}, 
Matplotlib \citep{Hunter_2007}, 
NumPy \citep{Harris_2020}, 
pandas \citep{pandasDT_2020, McKinney_2010},
SciPy \citep{Virtanen_2020}.
}

\bibliography{uno}{}

\begin{thebibliography}{}
\expandafter\ifx\csname natexlab\endcsname\relax\def\natexlab#1{#1}\fi
\providecommand{\url}[1]{\href{#1}{#1}}
\providecommand{\dodoi}[1]{doi:~\href{http://doi.org/#1}{\nolinkurl{#1}}}
\providecommand{\doeprint}[1]{\href{http://ascl.net/#1}{\nolinkurl{http://ascl.net/#1}}}
\providecommand{\doarXiv}[1]{\href{https://arxiv.org/abs/#1}{\nolinkurl{https://arxiv.org/abs/#1}}}

\bibitem[{{Alexandersen} {et~al.}(2016){Alexandersen}, {Gladman}, {Kavelaars},
  {Petit}, {Gwyn}, {Shankman}, \& {Pike}}]{Alexandersen_2016}
{Alexandersen}, M., {Gladman}, B., {Kavelaars}, J.~J., {et~al.} 2016, \aj, 152,
  111, \dodoi{10.3847/0004-6256/152/5/111}

\bibitem[{{Anaconda Software Distribution}(2020)}]{anaconda}
{Anaconda Software Distribution}. 2020, Anaconda [Computer software], Vers.
  2-2.4.0,  Anaconda Inc.
\newblock \url{https://docs.anaconda.com/}

\bibitem[{Anderson \& Darling(1952)}]{Anderson_1952}
Anderson, T.~W., \& Darling, D.~A. 1952, The Annals of Mathematical Statistics,
  23, 193 , \dodoi{10.1214/aoms/1177729437}

\bibitem[{{Bannister}(2020)}]{Bannister_2020}
{Bannister}, M.~T. 2020, in The Trans-Neptunian Solar System, ed.
  D.~{Prialnik}, M.~A. {Barucci}, \& L.~{Young}, 439--453,
  \dodoi{10.1016/B978-0-12-816490-7.00020-5}

\bibitem[{{Bannister} {et~al.}(2016){Bannister}, {Kavelaars}, {Petit},
  {Gladman}, {Gwyn}, {Chen}, {Volk}, {Alexandersen}, {Benecchi}, {Delsanti},
  {Fraser}, {Granvik}, {Grundy}, {Guilbert-Lepoutre}, {Hestroffer}, {Ip},
  {Jakubik}, {Jones}, {Kaib}, {Kavelaars}, {Lacerda}, {Lawler}, {Lehner},
  {Lin}, {Lister}, {Lykawka}, {Monty}, {Marsset}, {Murray-Clay}, {Noll},
  {Parker}, {Pike}, {Rousselot}, {Rusk}, {Schwamb}, {Shankman}, {Sicardy},
  {Vernazza}, \& {Wang}}]{Bannister_2016}
{Bannister}, M.~T., {Kavelaars}, J.~J., {Petit}, J.-M., {et~al.} 2016, \aj,
  152, 70, \dodoi{10.3847/0004-6256/152/3/70}

\bibitem[{{Bannister} {et~al.}(2018){Bannister}, {Gladman}, {Kavelaars},
  {Petit}, {Volk}, {Chen}, {Alexandersen}, {Gwyn}, {Schwamb}, {Ashton},
  {Benecchi}, {Cabral}, {Dawson}, {Delsanti}, {Fraser}, {Granvik},
  {Greenstreet}, {Guilbert-Lepoutre}, {Ip}, {Jakubik}, {Jones}, {Kaib},
  {Lacerda}, {Van Laerhoven}, {Lawler}, {Lehner}, {Lin}, {Lykawka}, {Marsset},
  {Murray-Clay}, {Pike}, {Rousselot}, {Shankman}, {Thirouin}, {Vernazza}, \&
  {Wang}}]{Bannister_2018}
{Bannister}, M.~T., {Gladman}, B.~J., {Kavelaars}, J.~J., {et~al.} 2018, \apjs,
  236, 18, \dodoi{10.3847/1538-4365/aab77a}

\bibitem[{{Bernardinelli} {et~al.}(2023){Bernardinelli}, {Napier},
  {Smotherman}, {Strauss}, {Trilling}, {Trujillo}, {Juric}, {Gerdes}, {Lin},
  {Chandler}, {Adams}, {Portillo}, {Stetzler}, {Holman}, {Payne}, {Simpson},
  {Fuentes}, {Oldroyd}, {Sheppard}, {Mommert}, {Markwardt}, {McNeill},
  {Schlichting}, {Ragozzine}, {Rivkin}, \& {Porter}}]{Bernardinelli_2023}
{Bernardinelli}, P., {Napier}, K., {Smotherman}, H., {et~al.} 2023, in American
  Astronomical Society Meeting Abstracts, Vol.~55, American Astronomical
  Society Meeting Abstracts, 136.05

\bibitem[{{Brunini} \& {Melita}(1998)}]{Brunini_1998}
{Brunini}, A., \& {Melita}, M.~D. 1998, \icarus, 135, 408,
  \dodoi{10.1006/icar.1998.5992}

\bibitem[{{Cabral} {et~al.}(2019){Cabral}, {Guilbert-Lepoutre}, {Fraser},
  {Marsset}, {Volk}, {Petit}, {Rousselot}, {Alexandersen}, {Bannister}, {Chen},
  {Gladman}, {Gwyn}, \& {Kavelaars}}]{Cabral_2019}
{Cabral}, N., {Guilbert-Lepoutre}, A., {Fraser}, W.~C., {et~al.} 2019, \aap,
  621, A102, \dodoi{10.1051/0004-6361/201834021}

\bibitem[{{Chang} {et~al.}(2021){Chang}, {Chen}, {Fraser}, {Yoshida}, {Lehner},
  {Wang}, {Kavelaars}, {Pike}, {Alexandersen}, {Ito}, {Choi}, {Granados
  Contreras}, {Jeongahn}, {Ji}, {Kim}, {Lawler}, {Li}, {Lin}, {Sofia Lykawka},
  {Moon}, {More}, {Mu{\~n}oz-Guti{\'e}rrez}, {Ohtsuki}, {Terai}, {Urakawa},
  {Zhang}, {Zhao}, {Zhou}, \& {Fossil Collaboration}}]{Chang_2021}
{Chang}, C.-K., {Chen}, Y.-T., {Fraser}, W.~C., {et~al.} 2021, \psj, 2, 191,
  \dodoi{10.3847/PSJ/ac13a4}

\bibitem[{Dirk(2021)}]{Dirk_2021}
Dirk, M.~E. 2021, Journal of the American Academy of Dermatology,
  \dodoi{10.1016/j.jaad.2021.06.845}

\bibitem[{{Duffard} {et~al.}(2014){Duffard}, {Pinilla-Alonso}, {Santos-Sanz},
  {Vilenius}, {Ortiz}, {Mueller}, {Fornasier}, {Lellouch}, {Mommert}, {Pal},
  {Kiss}, {Mueller}, {Stansberry}, {Delsanti}, {Peixinho}, \&
  {Trilling}}]{Duffard_2014}
{Duffard}, R., {Pinilla-Alonso}, N., {Santos-Sanz}, P., {et~al.} 2014, \aap,
  564, A92, \dodoi{10.1051/0004-6361/201322377}

\bibitem[{Dvorak {et~al.}(2007)Dvorak, Schwarz, Süli, \&
  Kotoulas}]{Dvorak_2007}
Dvorak, R., Schwarz, R., Süli, {\'A}., \& Kotoulas, T. 2007, Monthly Notices
  of the Royal Astronomical Society, 382, 1324,
  \dodoi{10.1111/j.1365-2966.2007.12480.x}

\bibitem[{{Fern{\'a}ndez} {et~al.}(2018){Fern{\'a}ndez}, {Helal}, \&
  {Gallardo}}]{Fernandez_2018}
{Fern{\'a}ndez}, J.~A., {Helal}, M., \& {Gallardo}, T. 2018, \planss, 158, 6,
  \dodoi{10.1016/j.pss.2018.05.013}

\bibitem[{{Fraser} {et~al.}(2022){Fraser}, {Lawler}, {Ashton}, {Chen}, {Huang},
  {Gladman}, {Kavelaars}, {Petit}, {Peltier}, {Pike}, {Alexandersen},
  {Hestroffer}, {Noyelles}, {Chang}, {Wang}, {Connolly}, {Kalmbach}, {Eduardo},
  {Juric}, \& {Gwyn}}]{Fraser_2022}
{Fraser}, W., {Lawler}, S., {Ashton}, E., {et~al.} 2022, in AAS/Division for
  Planetary Sciences Meeting Abstracts, Vol.~54, AAS/Division for Planetary
  Sciences Meeting Abstracts, 414.01

\bibitem[{{Gladman} \& {Duncan}(1990)}]{Gladman_1990}
{Gladman}, B., \& {Duncan}, M. 1990, \aj, 100, 1680, \dodoi{10.1086/115628}

\bibitem[{{Gomes}(2021)}]{Gomes_2021}
{Gomes}, R. 2021, \icarus, 357, 114121, \dodoi{10.1016/j.icarus.2020.114121}

\bibitem[{{Grav} {et~al.}(2011){Grav}, {Jedicke}, {Denneau}, {Chesley},
  {Holman}, \& {Spahr}}]{Grav_2011}
{Grav}, T., {Jedicke}, R., {Denneau}, L., {et~al.} 2011, \pasp, 123, 423,
  \dodoi{10.1086/659833}

\bibitem[{Harris {et~al.}(2020)Harris, Millman, van~der Walt, Gommers,
  Virtanen, Cournapeau, Wieser, Taylor, Berg, Smith, Kern, Picus, Hoyer, van
  Kerkwijk, Brett, Haldane, del R{\'{i}}o, Wiebe, Peterson,
  G{\'{e}}rard-Marchant, Sheppard, Reddy, Weckesser, Abbasi, Gohlke, \&
  Oliphant}]{Harris_2020}
Harris, C.~R., Millman, K.~J., van~der Walt, S.~J., {et~al.} 2020, Nature, 585,
  357, \dodoi{10.1038/s41586-020-2649-2}

\bibitem[{{Holman}(1997)}]{Holman_1997}
{Holman}, M.~J. 1997, \nat, 387, 785

\bibitem[{{Holman} \& {Wisdom}(1993)}]{Holman_1993}
{Holman}, M.~J., \& {Wisdom}, J. 1993, \aj, 105, 1987, \dodoi{10.1086/116574}

\bibitem[{{Horner} {et~al.}(2004){Horner}, {Evans}, \& {Bailey}}]{Horner_2004}
{Horner}, J., {Evans}, N.~W., \& {Bailey}, M.~E. 2004, \mnras, 354, 798,
  \dodoi{10.1111/j.1365-2966.2004.08240.x}

\bibitem[{Hunter(2007)}]{Hunter_2007}
Hunter, J.~D. 2007, Computing in Science \& Engineering, 9, 90,
  \dodoi{10.1109/MCSE.2007.55}

\bibitem[{{Kavelaars} {et~al.}(2008){Kavelaars}, {Jones}, {Gladman}, {Parker},
  \& {Petit}}]{Kavelaars_2008}
{Kavelaars}, J., {Jones}, L., {Gladman}, B., {Parker}, J.~W., \& {Petit}, J.~M.
  2008, in The Solar System Beyond Neptune, ed. M.~A. {Barucci},
  H.~{Boehnhardt}, D.~P. {Cruikshank}, A.~{Morbidelli}, \& R.~{Dotson}, 59

\bibitem[{{Lawler} {et~al.}(2018{\natexlab{a}}){Lawler}, {Kavelaars},
  {Alexandersen}, {Bannister}, {Gladman}, {Petit}, \&
  {Shankman}}]{Lawler_2018_O10}
{Lawler}, S.~M., {Kavelaars}, J.~J., {Alexandersen}, M., {et~al.}
  2018{\natexlab{a}}, Frontiers in Astronomy and Space Sciences, 5, 14,
  \dodoi{10.3389/fspas.2018.00014}

\bibitem[{{Lawler} {et~al.}(2018{\natexlab{b}}){Lawler}, {Shankman},
  {Kavelaars}, {Alexandersen}, {Bannister}, {Chen}, {Gladman}, {Fraser},
  {Gwyn}, {Kaib}, {Petit}, \& {Volk}}]{Lawler_2018_O8}
{Lawler}, S.~M., {Shankman}, C., {Kavelaars}, J.~J., {et~al.}
  2018{\natexlab{b}}, \aj, 155, 197, \dodoi{10.3847/1538-3881/aab8ff}

\bibitem[{{Lin} {et~al.}(2021){Lin}, {Chen}, {Volk}, {Gladman}, {Murray-Clay},
  {Alexandersen}, {Bannister}, {Lawler}, {Ip}, {Lykawka}, {Kavelaars}, {Gwyn},
  \& {Petit}}]{Lin_2021}
{Lin}, H.~W., {Chen}, Y.-T., {Volk}, K., {et~al.} 2021, \icarus, 361, 114391,
  \dodoi{10.1016/j.icarus.2021.114391}

\bibitem[{{LSST Science Collaboration} {et~al.}(2009){LSST Science
  Collaboration}, {Abell}, {Allison}, {Anderson}, {Andrew}, {Angel}, {Armus},
  {Arnett}, {Asztalos}, {Axelrod}, {Bailey}, {Ballantyne}, {Bankert},
  {Barkhouse}, {Barr}, {Barrientos}, {Barth}, {Bartlett}, {Becker}, {Becla},
  {Beers}, {Bernstein}, {Biswas}, {Blanton}, {Bloom}, {Bochanski}, {Boeshaar},
  {Borne}, {Bradac}, {Brandt}, {Bridge}, {Brown}, {Brunner}, {Bullock},
  {Burgasser}, {Burge}, {Burke}, {Cargile}, {Chandrasekharan}, {Chartas},
  {Chesley}, {Chu}, {Cinabro}, {Claire}, {Claver}, {Clowe}, {Connolly}, {Cook},
  {Cooke}, {Cooray}, {Covey}, {Culliton}, {de Jong}, {de Vries}, {Debattista},
  {Delgado}, {Dell'Antonio}, {Dhital}, {Di Stefano}, {Dickinson}, {Dilday},
  {Djorgovski}, {Dobler}, {Donalek}, {Dubois-Felsmann}, {Durech},
  {Eliasdottir}, {Eracleous}, {Eyer}, {Falco}, {Fan}, {Fassnacht}, {Ferguson},
  {Fernandez}, {Fields}, {Finkbeiner}, {Figueroa}, {Fox}, {Francke}, {Frank},
  {Frieman}, {Fromenteau}, {Furqan}, {Galaz}, {Gal-Yam}, {Garnavich},
  {Gawiser}, {Geary}, {Gee}, {Gibson}, {Gilmore}, {Grace}, {Green}, {Gressler},
  {Grillmair}, {Habib}, {Haggerty}, {Hamuy}, {Harris}, {Hawley}, {Heavens},
  {Hebb}, {Henry}, {Hileman}, {Hilton}, {Hoadley}, {Holberg}, {Holman},
  {Howell}, {Infante}, {Ivezic}, {Jacoby}, {Jain}, {R}, {Jedicke}, {Jee},
  {Garrett Jernigan}, {Jha}, {Johnston}, {Jones}, {Juric}, {Kaasalainen},
  {Styliani}, {Kafka}, {Kahn}, {Kaib}, {Kalirai}, {Kantor}, {Kasliwal},
  {Keeton}, {Kessler}, {Knezevic}, {Kowalski}, {Krabbendam}, {Krughoff},
  {Kulkarni}, {Kuhlman}, {Lacy}, {Lepine}, {Liang}, {Lien}, {Lira}, {Long},
  {Lorenz}, {Lotz}, {Lupton}, {Lutz}, {Macri}, {Mahabal}, {Mandelbaum},
  {Marshall}, {May}, {McGehee}, {Meadows}, {Meert}, {Milani}, {Miller},
  {Miller}, {Mills}, {Minniti}, {Monet}, {Mukadam}, {Nakar}, {Neill}, {Newman},
  {Nikolaev}, {Nordby}, {O'Connor}, {Oguri}, {Oliver}, {Olivier}, {Olsen},
  {Olsen}, {Olszewski}, {Oluseyi}, {Padilla}, {Parker}, {Pepper}, {Peterson},
  {Petry}, {Pinto}, {Pizagno}, {Popescu}, {Prsa}, {Radcka}, {Raddick},
  {Rasmussen}, {Rau}, {Rho}, {Rhoads}, {Richards}, {Ridgway}, {Robertson},
  {Roskar}, {Saha}, {Sarajedini}, {Scannapieco}, {Schalk}, {Schindler},
  {Schmidt}, {Schmidt}, {Schneider}, {Schumacher}, {Scranton}, {Sebag},
  {Seppala}, {Shemmer}, {Simon}, {Sivertz}, {Smith}, {Allyn Smith}, {Smith},
  {Spitz}, {Stanford}, {Stassun}, {Strader}, {Strauss}, {Stubbs}, {Sweeney},
  {Szalay}, {Szkody}, {Takada}, {Thorman}, {Trilling}, {Trimble}, {Tyson}, {Van
  Berg}, {Vanden Berk}, {VanderPlas}, {Verde}, {Vrsnak}, {Walkowicz},
  {Wandelt}, {Wang}, {Wang}, {Warner}, {Wechsler}, {West}, {Wiecha},
  {Williams}, {Willman}, {Wittman}, {Wolff}, {Wood-Vasey}, {Wozniak}, {Young},
  {Zentner}, \& {Zhan}}]{LSSTSC_2009}
{LSST Science Collaboration}, {Abell}, P.~A., {Allison}, J., {et~al.} 2009,
  arXiv e-prints, arXiv:0912.0201.
\newblock \doarXiv{0912.0201}

\bibitem[{{Lykawka}(2012)}]{Lykawka_2012}
{Lykawka}, P.~S. 2012, Monographs on Environment, Earth and Planets, 1, 121,
  \dodoi{10.5047/meep.2012.00103.0121}

\bibitem[{McKinney(2010)}]{McKinney_2010}
McKinney, W. 2010, in {P}roceedings of the 9th {P}ython in {S}cience
  {C}onference, ed. S.~{van der Walt} \& J.~{Millman}, 56 -- 61,
  \dodoi{10.25080/Majora-92bf1922-00a}

\bibitem[{{Morbidelli} \& {Nesvorn{\'y}}(2020)}]{Morbidelli_2020}
{Morbidelli}, A., \& {Nesvorn{\'y}}, D. 2020, in The Trans-Neptunian Solar
  System, ed. D.~{Prialnik}, M.~A. {Barucci}, \& L.~{Young}, 25--59,
  \dodoi{10.1016/B978-0-12-816490-7.00002-3}

\bibitem[{{Nesvorn{\'y}} \& {Morbidelli}(2012)}]{Nesvorny_2012}
{Nesvorn{\'y}}, D., \& {Morbidelli}, A. 2012, \aj, 144, 117,
  \dodoi{10.1088/0004-6256/144/4/117}

\bibitem[{{Nesvorn{\'y}} {et~al.}(2013){Nesvorn{\'y}}, {Vokrouhlick{\'y}}, \&
  {Morbidelli}}]{Nesvorny_2013}
{Nesvorn{\'y}}, D., {Vokrouhlick{\'y}}, D., \& {Morbidelli}, A. 2013, \apj,
  768, 45, \dodoi{10.1088/0004-637X/768/1/45}

\bibitem[{{Nesvorn{\'y}} {et~al.}(2020){Nesvorn{\'y}}, {Vokrouhlick{\'y}},
  {Alexandersen}, {Bannister}, {Buchanan}, {Chen}, {Gladman}, {Gwyn},
  {Kavelaars}, {Petit}, {Schwamb}, \& {Volk}}]{Nesvorny_2020}
{Nesvorn{\'y}}, D., {Vokrouhlick{\'y}}, D., {Alexandersen}, M., {et~al.} 2020,
  \aj, 160, 46, \dodoi{10.3847/1538-3881/ab98fb}

\bibitem[{pandas~development team(2020)}]{pandasDT_2020}
pandas~development team, T. 2020, pandas-dev/pandas: Pandas, latest,  Zenodo,
  \dodoi{10.5281/zenodo.3509134}

\bibitem[{{Peixinho} {et~al.}(2015){Peixinho}, {Delsanti}, \&
  {Doressoundiram}}]{Peixinho_2015}
{Peixinho}, N., {Delsanti}, A., \& {Doressoundiram}, A. 2015, \aap, 577, A35,
  \dodoi{10.1051/0004-6361/201425436}

\bibitem[{{Petit} {et~al.}(2011){Petit}, {Kavelaars}, {Gladman}, {Jones},
  {Parker}, {Van Laerhoven}, {Nicholson}, {Mars}, {Rousselot}, {Mousis},
  {Marsden}, {Bieryla}, {Taylor}, {Ashby}, {Benavidez}, {Campo Bagatin}, \&
  {Bernabeu}}]{Petit_2011}
{Petit}, J.~M., {Kavelaars}, J.~J., {Gladman}, B.~J., {et~al.} 2011, \aj, 142,
  131, \dodoi{10.1088/0004-6256/142/4/131}

\bibitem[{{Petit} {et~al.}(2017){Petit}, {Kavelaars}, {Gladman}, {Jones},
  {Parker}, {Bieryla}, {Van Laerhoven}, {Pike}, {Nicholson}, {Ashby}, \&
  {Lawler}}]{Petit_2017}
---. 2017, \aj, 153, 236, \dodoi{10.3847/1538-3881/aa6aa5}

\bibitem[{{Schwamb} {et~al.}(2023){Schwamb}, {Jones}, {Yoachim}, {Volk},
  {Dorsey}, {Opitom}, {Greenstreet}, {Lister}, {Snodgrass}, {Bolin}, {Inno},
  {Bannister}, {Eggl}, {Solontoi}, {Kelley}, {Juri{\'c}}, {Lin}, {Ragozzine},
  {Bernardinelli}, {Chesley}, {Daylan}, {{\v{D}}urech}, {Fraser}, {Granvik},
  {Knight}, {Lisse}, {Malhotra}, {Oldroyd}, {Thirouin}, \& {Ye}}]{Schwamb_2023}
{Schwamb}, M.~E., {Jones}, R.~L., {Yoachim}, P., {et~al.} 2023, arXiv e-prints,
  arXiv:2303.02355, \dodoi{10.48550/arXiv.2303.02355}

\bibitem[{{The Rubin Observatory Survey Cadence Optimization
  Committee}(2023)}]{ROSCOC_2023}
{The Rubin Observatory Survey Cadence Optimization Committee}. 2023, Survey
  Cadence Optimization Committee’s Phase 2 Recommendations.
\newblock \url{https://pstn-055.lsst.io/}

\bibitem[{{Tiscareno} \& {Malhotra}(2003)}]{Tiscareno_2003}
{Tiscareno}, M.~S., \& {Malhotra}, R. 2003, \aj, 126, 3122,
  \dodoi{10.1086/379554}

\bibitem[{Virtanen {et~al.}(2020)Virtanen, Gommers, Oliphant, Haberland, Reddy,
  Cournapeau, Burovski, Peterson, Weckesser, Bright, {van der Walt}, Brett,
  Wilson, Millman, Mayorov, Nelson, Jones, Kern, Larson, Carey, Polat, Feng,
  Moore, {VanderPlas}, Laxalde, Perktold, Cimrman, Henriksen, Quintero, Harris,
  Archibald, Ribeiro, Pedregosa, {van Mulbregt}, \& {SciPy 1.0
  Contributors}}]{Virtanen_2020}
Virtanen, P., Gommers, R., Oliphant, T.~E., {et~al.} 2020, Nature Methods, 17,
  261, \dodoi{10.1038/s41592-019-0686-2}

\bibitem[{{Zhang} \& {Gladman}(2022)}]{Zhang_2022}
{Zhang}, K., \& {Gladman}, B.~J. 2022, \na, 90, 101659,
  \dodoi{10.1016/j.newast.2021.101659}

\end{thebibliography}
\bibliographystyle{aasjournal}

\end{document}